\documentclass[journal]{IEEEtran}

\usepackage{amsmath,cite}
\usepackage{graphicx}  
\usepackage{subfigure}
\usepackage{color}
\usepackage[linesnumbered,ruled]{algorithm2e}
\usepackage{subfig}
\usepackage{tabularx}
\usepackage{rotating}
\usepackage{tikz}
\usepackage{url}
\usepackage{changes}

\usepackage{times}
\usepackage{ifpdf}
\usepackage[english]{babel}

\usepackage{booktabs}
\usepackage{colortbl}

\definecolor{cadmiumgreen}{rgb}{0.0, 0.42, 0.24}
\definecolor{darkblue}{rgb}{0.0, 0.0, 0.7}
\definecolor{tablecaption}{rgb}{0.95, 0.95, 0.95}

\newcommand{\cmmnt}[1]{}

\ifCLASSINFOpdf
\else
\fi

\begin{document}

\title{Creating A Multi-track Classical Music Performance Dataset for Multi-modal Music Analysis: Challenges, Insights, and Applications}

\author{
        Bochen Li$^{\star}$,
        Xinzhao Liu$^{\star}$,
        Karthik Dinesh,
        Zhiyao Duan,~\IEEEmembership{Member,~IEEE,}
        and~Gaurav~Sharma,~\IEEEmembership{Fellow,~IEEE}
        
 \thanks{$^{\star}$B. Li and X. Liu made equal contribution to this paper.}
\thanks{X. Liu was with the Department of Electrical and Computer Engineering, University of Rochester, Rochester, NY 14627, USA. She is now with Listent American Corp. E-mail: xinzhao.liu@rochester.edu}%
\thanks{B. Li, K. Dinesh, Z. Duan, and G. Sharma are with the Department of Electrical and Computer Engineering, University of Rochester, Rochester, NY, 14627, USA. E-mails: \{bochen.li, kdinesh, zhiyao.duan, gaurav.sharma\}@rochester.edu.} 
}

\maketitle

\begin{abstract}

We introduce a dataset for facilitating audio-visual analysis of music performances. The dataset comprises 44 simple multi-instrument classical music pieces assembled from coordinated but separately recorded performances of individual tracks. For each piece, we provide the musical score in MIDI format, the audio recordings of the individual tracks, the audio and video recording of the assembled mixture, and ground-truth annotation files including frame-level and note-level transcriptions. We describe our methodology for the creation of the dataset, particularly highlighting our approaches for addressing the challenges involved in maintaining synchronization and expressiveness. We demonstrate the high quality of synchronization achieved with our proposed approach by comparing the dataset with existing widely-used music audio datasets. 

We anticipate that the dataset will be useful for the development and evaluation of existing music information retrieval (MIR) tasks, as well as for novel multi-modal tasks. We benchmark two existing MIR tasks (multi-pitch analysis and score-informed source separation) on the dataset and compare with other existing music audio datasets. Additionally, we consider two novel multi-modal MIR tasks (visually informed multi-pitch analysis and polyphonic vibrato analysis) enabled by the dataset and provide evaluation measures and baseline systems for future comparisons (from our recent work). Finally, we propose several emerging research directions that the dataset enables.
 \end{abstract}


\IEEEpeerreviewmaketitle

\section{Introduction}
\label{sec:intro}
\IEEEPARstart{M}{usic} performance is a multi-modal art form. For thousands of years, people have enjoyed music performances at live concerts through both hearing and sight. The development of recording technologies, starting with Thomas Edison's invention of the phonograph in 1877, has extended music enjoyment beyond live concerts. For a long time, the majority of music recordings were distributed through various kinds of media that carry only the audio, such as vinyl records, cassettes, CDs, and mp3 files. As such, existing research on music analysis, processing, and retrieval focuses on the audio modality, while the visual component was largely forgotten.

About a decade ago, with the rapid expansion of digital storage and internet bandwidth, video streaming services like YouTube gained popularity, which again significantly influenced the way people enjoy music. In addition to listening to the sound, audiences also want to watch the performance. In 2014, “music” was the most searched topic on YouTube, and 38.4\% of YouTube views were from music videos \cite{smithdata}. The visual modality plays an important role in music performances. Guitar players learn new songs by watching how others play online. Concert attendees move their gazes to the soloist in a jazz concert. In fact, researchers have found that “the visual component is not just a marginal phenomenon in music perception, but an important factor in the communication of meanings” \cite{platz2012eye}. Even for prestigious classical music competitions, researchers have found that visually perceived elements of the performance, such as gesture, motion, and facial expressions of the performer, affect the judge's (experts or novice alike) evaluations, even more significantly than the sound \cite{tsay2013sight}.

Music Information Retrieval (MIR) research, which traditionally focused on audio and symbolic modalities (e.g., musical scores), started to pay attention to other modalities in recent years. For example, players' motion data captured from sensors (MoCap) have been used to analyze players' activities \cite{caramiaux2012segmenting} and enhance source separation quality \cite{parekh2017motion}. However, such data is not easy to obtain from natural music performances because the methodology requires specialized motion capture sensors.

The visual modality is much more natural, and when available, it can be very helpful for solving many MIR tasks that are challenging using an audio-only approach. Zhang et al. \cite{zhang2007visual} introduced a method to transcribe solo violin performances by tracking the violin strings and fingers from the visual scene. Similarly, remarkable success has been demonstrated for music transcription using visual techniques for piano~\cite{akbari2015real}, guitar \cite{paleari2008multimodal} and drums \cite{gillet2005automatic}. Other MIR tasks include onset detection \cite{bazzica2017vision} and vibrato analysis \cite{li2017video}.
Note that the benefits of incorporating visual information in the analysis of audio are especially pronounced for highly polyphonic, multi-instrument performances, because the visual activity of each player is usually directly observable (barring occlusions), whereas the polyphony makes it difficult to unambiguously associate audio components with each player. Dinesh et al. \cite{dinesh2017visually} proposed to detect play/non-play activity for each player in a string ensemble to achieve improved multi-pitch estimation and streaming results than audio-based methods. A similar idea is applied on different instrument groups among symphony orchestras to achieve performance-score alignment \cite{bazzica2014exploiting}. Additionally, audio-visual analysis opens up new frontiers of MIR research. Researchers have proposed systems to analyze the fingering of guitarists \cite{burns2006visual, kerdvibulvech2007vision, radicioni2004segmentation, scarr2010retrieval} and pianists \cite{gorodnichy2006detection, oka2013marker}, the baton trajectories of the conductors \cite{murphy2003tracking}, the audio-visual source association in multimedia music \cite{li2017see}, and the interaction modes between players and instruments \cite{yao2013discovering}.

Despite the increased recent interest, progress in jointly using audio and visual modalities for the analysis of music performances has been rather slow. One of the main reasons, we argue, is the lack of datasets. Although music takes a large share among all kinds of multimedia data, music datasets are scarce. This is because a music dataset should contain not only music recordings but also ground-truth annotations (e.g., note/beat/chord transcriptions, performance-score alignments) to enable supervised machine learning and the evaluation of proposed methods. Due to the temporal and polyphonic nature of music, the annotation process is very time consuming and often requires significant musical expertise. Furthermore, for some research problems such as source separation, isolated recordings of different sound sources (e.g., musical instruments) are also needed for ground truth verification. When creating such a dataset, if each source is recorded in isolation, it is a challenging task to ensure that different sources are well tuned and properly synchronized.

In this paper, we present the University of Rochester Multi-modal Music Performance (URMP) dataset. This dataset covers 44 classical chamber music pieces ranging from duets to quintets. For each included piece, the URMP dataset provides the musical score, the audio recordings of the individual tracks, the audio and video recordings of the assembled mixture, and ground-truth annotation files including frame-level and note-level transcriptions. In creating the URMP dataset, a key challenge we encountered and overcame was the synchronization of individually recorded instrumental sources of a piece while maintaining the expressiveness seen in professional chamber music performances. We present our attempts and reflections on addressing this challenge, and hope that this will shed light on similar issues for future dataset creation efforts. We also conduct objective and subjective evaluations on the synchronization quality and compare it with two widely used datasets. As the first audio-visual multi-instrument multi-track music performance dataset, we anticipate that it will be valuable for MIR research. Therefore, we benchmark URMP with existing widely used music audio datasets on important existing tasks. We also highlight our previous work on URMP to define two novel multi-modal MIR tasks by proposing evaluation measures and providing baseline systems for comparison. We further propose several emerging novel research directions that URMP may support.

In the rest of the paper, in Section \ref{sec:review}, we first review existing music performance datasets for MIR tasks and especially highlight the challenges involved in creating multi-track datasets. Then, in Section \ref{sec:attempts}, we describe our different attempts aimed at overcoming these challenges while recording the URMP dataset and, in Section \ref{sec:procedure}, elaborate on the approach adopted. In Section \ref{sec:dataset}, we describe the content of the URMP dataset and analyze the quality of the dataset. In Section \ref{sec:applications}, we compare the URMP dataset with other existing music audio datasets by benchmarking two pre-existing audio-only MIR tasks on the datasets and also mention several novel multi-modal MIR tasks enabled by the multi-modal data in URMP. Finally, we conclude the paper in Section \ref{sec:conclusions}.

\section{Review of Music Performance Datasets}
\label{sec:review}


\begin{table*}[h]
\renewcommand{\arraystretch}{1.2}
\centering
\small
\begin{tabular}{p{3cm}cccp{6cm}}

\toprule[1.5pt]
\multicolumn{1}{c}{ \textbf{Name} }    & 
\textbf{ Instrument/Genre } & \textbf{ \# Pieces } & \textbf{ Total Duration }  & \multicolumn{1}{c}{ \textbf{Content} } \\ 

\midrule[1.5pt]

\hline
\rowcolor{tablecaption}
\multicolumn{5}{c}{ 
\textit{Audio-modality, Single-track}
}\\

MAPS \cite{emiya2010multipitch}			& Piano 	& 270 	& 18.6 h  	& Audio, Note annotation \\

LabROSA \cite{poliner2007discriminative} & Piano 	& 130 	& 2.7 h 	& Audio, Note annotation \\

Score-informed Piano Transcription \cite{benetos2012score}  & Piano 	& 7 	& 6.4 m & Audio, Note annotation, Performance error annotation \\

RWC \cite{goto2002rwc} (Subset) & Multi-instrument & 100 & 9.2 h & Audio, Note annotation \\

Su et al. \cite{su2015escaping} 	& Multi-instrument 	& 10 	& 5 m & Audio, Note annotation \\

\hline
\rowcolor{tablecaption}
\multicolumn{5}{c}{ 
\textit{Audio-modality, Multi-track}
}\\

MedleyDB \cite{bittner2014medleydb} 	& Multi-genre	& 122	& 7.3 h & Audio, Pitch contour, Instrument activity, Genre label \\

SSMD \cite{hargreaves2012structural}	& Songs			& 104	& 6.8 h	& Audio, Structure annotation \\

MASS \cite{MTGMASSdb}		& Songs, Multi-genre				& 6		& 4.8 m		& Audio, Lyrics \\

Mixploration \cite{cartwright2014mixploration}	& Multi-genre			& 12     & 4.9 m	& Audio \\

iKala \cite{chan2015vocal}		& Songs		& 252	& 2.1 h	& Audio, Lyrics, Pitch contour \\

WWQ \cite{bay2009evaluation}			& Multi-instrument	& 1	& 9 m$^\star$	& Audio, Note annotation \\

TRIOS \cite{fritsch2013score}		& Multi-instrument	& 5	&3.2 m	& Audio, Note annotation \\

Bach10 \cite{duan2010multiple}		& Multi-instrument	& 10	& 5.5 m	& Audio, Note annotation, Pitch contour \\

PHENICX-Anechoic \cite{patynen2008anechoic, miron2016score}		& Multi-instrument	& 4	& 10.6 m	& Audio, Note annotation \\

\hline
\rowcolor{tablecaption}
\multicolumn{5}{c}{ 
\textit{Multi-modality, Single-track}
}\\

Multi-modal Guitar \cite{perezestimation}	& Guitar	& 10	& 10 m	& Audio, Video \\

C4S	\cite{bazzica2017vision}				& Clarinet	& 54	& 4.5 h	& Audio, Video, Visual annotation \\

\hline
\rowcolor{tablecaption}
\multicolumn{5}{c}{ 
\textit{Multi-modality, Multi-track}
}\\

ENST-Drums \cite{gillet2006enst}& Drum kit	& N/A	& 3.75 h	& Audio, Video (multi-camera views) \\

Abe{\ss}er et al. \cite{abesser2011modeling}	& Guitar, Drum, Bass	& N/A	& 1.2 h	& Audio, Video (multi-camera views) \\

EEP \cite{marchini2014sense}			& String quartet	& 23		& N/A		& Audio, Note annotation, Bow MoCap data \\

URMP		& Multi-instrument	& 44		& 1.3 h		& Audio, Video, Note annotation, Pitch contour\\

\hline

\end{tabular}

\caption{Summary of several commonly used music performance datasets for music transcription, source separation, audio-score alignment, and multi-modal music analysis. $\star$): Only 54 seconds are publicly available.} \label{tab:ReviewDataset}
\end{table*}

Music performance datasets are not easy to create because recording music performances and annotating them with ground-truth labels (e.g., pitch, chord, structure, and mood) require musical expertise and are very time consuming. Commercial recordings can generally not be used due to copyright issues. 
Recording music performances in research labs is subject to the availability of musicians and recording facilities. Also, when different instruments are recorded in isolation (for evaluating musical source separation), we need to ensure proper methods for synchronization.
The annotation process often requires experienced musicians to listen through the musical recording multiple times. It is especially difficult when the annotations are numerical and at a temporal resolution on the order of milliseconds (for evaluating pitch transcription, audio-score alignment, etc.). As a result, music performance datasets are scarce and their sizes are relatively small. 

\subsection{Existing Datasets}

In this section, we briefly review several commonly-used music performance datasets that are closely related to the URMP dataset, which can support some MIR tasks like music transcription, source separation, audio-score alignment, etc. A summary of these datasets is provided in Table \ref{tab:ReviewDataset}. Most of the datasets contain only audio, and only six are multi-modal.

The first group of datasets are single-track polyphonic recordings with MIDI transcriptions for music transcription research. While recording this type of music is straightforward, obtaining the ground-truth transcription is not. A large portion of existing single-track datasets focus on piano music \cite{emiya2010multipitch,poliner2007discriminative,benetos2012score}, where the transcription can be obtained automatically: a pianist plays on a MIDI keyboard to generate a MIDI performance with precise note timings and dynamics; the MIDI file is then fed to a reproducing piano (e.g., Yamaha Disklavier\footnote{\url{http://www.disklavier.com}}) to render acoustical recordings.
The MIDI file naturally serves as the ground-truth transcription. For other instruments that do not have the MIDI-driven sound reproducing systems, manual annotation of ground-truth transcription is the most accurate approach, which, however, is notoriously labor intensive. To address this issue, the RWC dataset \cite{goto2002rwc} (classical and jazz subset) aligns a reference MIDI score to the audio performance in a semi-automatic fashion, and uses the aligned MIDI score as the transcription. The dataset proposed by Su et al. \cite{su2015escaping} uses a different approach, where a professional pianist was employed to follow and play the music on an electric piano to generate well-aligned ground-truth transcriptions. 

The second group of datasets are multi-track recordings, where each instrumental source is on one track. A multi-track dataset generally has the merit of better versatility and scalability. First, it can support more MIR tasks (e.g., source separation). Second, it significantly reduces the annotation complexity, from
polyphonic to monophonic music. With a robust monophonic pitch analysis tool,  fine-grained annotations (e.g., pitch height in musical cents for each time frame) can be acquired with less labor. Third, a large variety of music excerpts can be reproduced by mixing the monophonic tracks of the same piece with different combinations. The difficulty in creating multi-track datasets is during the recording process, which will be discussed in the Section \ref{sec:review:challenges}.

The largest multi-track music dataset is MedleyDB \cite{bittner2014medleydb}. It contains multi-track audio recordings of 122 pieces with various styles together with the melody pitch contour and instrument activity annotations. The second largest dataset is the Structural Segmentation Multitrack Dataset (SSMD) \cite{hargreaves2012structural}, which contains multi-track audio recordings of 104 rock and pop songs, together with structural segmentation annotations. Most recordings of MedleyDB and SSMD are from third-party musical organizations (e.g., commercial or non-profit websites, recording studios). This lessens the burden of recoding by the researchers themselves. The other multi-track datasets are of a much smaller scale. The MASS dataset \cite{MTGMASSdb} contains several raw and effects-processed multi-track audio recordings. Mixploration dataset \cite{cartwright2014mixploration} contains 3 raw multi-track audio recordings together with a number of mixing parameters. 
The iKala dataset \cite{chan2015vocal} contains the vocal melody and the accompaniment part of 252 pop songs in separate tracks. The Wood Wind Quintet (WWQ) dataset  \cite{bay2009evaluation} contains individual recordings of 1 classical quintet. The original 9-min recording serves as an internal benchmark for the MIREX\footnote{\url{http://www.music-ir.org/mirex/wiki/MIREX_HOME}} Multi-F0 Estimation \& Tracking task since 2007; only a 54-second excerpt is publicly available. The TRIOS dataset \cite{fritsch2013score} contains 5 multi-track recordings of musical trios together with their MIDI transcriptions. The Bach10 dataset \cite{duan2010multiple} contains 10 multi-track instrumental recordings of \textit{J.S. Bach} four-part chorales, together with the pitch and note transcriptions and the ground-truth audio-score alignment. The PHENICX-Anechoic \cite{miron2016score} provides the denoised recordings and note annotations for the Aalto Anechoic Orchestral Database \cite{patynen2008anechoic}, which contains four symphony pieces, each one has 8-10 instrumental parts and each part was recorded in isolation using multiple microphones.

Existing multi-modal musical datasets include the Multi-modal Guitar dataset \cite{perezestimation}, the Clarinetists for Science (C4S) dataset \cite{bazzica2017vision}, the ENST-Drums dataset \cite{gillet2006enst}, the Abe{\ss}er et al. \cite{abesser2011modeling}, and the Ensemble Expressive Performance (EEP) dataset \cite{marchini2014sense}. The first two are single-track datasets. The Multi-modal Guitar dataset contains 10 audio-visual recordings of guitar performances. The audio was recorded using a contact microphone to capture the vibration of the guitar body and to attenuate the effects of room acoustics and sound radiation. The video was recorded using a high-speed camera with markers attached on joints of the player's hands and the guitar body to facilitate hand and instrument tracking. The C4S dataset consists of 54 videos from 9 clarinetists, each performing 3 different classical music pieces twice. Visual annotations are also provided including the coordinates of the face, mouth, left hand, right hand, and the clarinet. The latter three are multi-track datasets. The ENST-Drums contains mixed stereo audio tracks, audio and video recordings of each instrument of a drum kit playing different sequences. All instruments were recorded simultaneously using 8 microphones and 2 cameras. Similarly, instruments in \cite{abesser2011modeling} were also recorded simultaneously. Since different instruments were not recorded in isolation, there is some sound leakage across instrumental tracks. In EEP, each instrument was recorded using a contact microphone; while sound leakage is greatly reduced, the acoustic properties can be very different from using a near-field microphone. There are several other video datasets that contain a subset of music performance such as FCVID \cite{jiangfcvid}, YouTube-8M \cite{abu2016youtube}, Google AudioSet \cite{gemmeke2017audio}, etc. We do not include them in Table \ref{tab:ReviewDataset} since no content-level annotations are provided, which limits applications in MIR tasks.

\subsection{Synchronization Challenges in Creating Multi-track Datasets}
\label{sec:review:challenges}
The coordination between simultaneous sound sources differentiates music from general polyphonic acoustic scenes. One important aspect of this coordination is synchronization, which is typically accomplished in real-world music performances by players rehearsing together prior to a performance. During the performance, players also rely on auditory and visual cues to adjust their speed to other players. For large ensembles such as a symphony orchestra, a conductor sets the synchronization.

In order to have a music performance dataset simulate real-world scenarios, good synchronization between different instrumental parts is desired. However, creating a multi-track dataset without leakage across different tracks is challenging. To eliminate leakage, different instruments need to be recorded separately, which makes it difficult to achieve synchronization because players cannot rely on interactions with other players to adjust their timing. In this subsection, we review existing approaches that researchers have explored for ensuring synchronization when recording multi-track datasets.

For SSMD, MASS, Mixploration, and a large portion of MelodyDB, recordings were obtained from professional musical organizations and recording studios instead of being recorded in a laboratory setting. The pieces are also mostly rock and pop songs, which have a steady tempo, making synchronization easier. In fact, in the music production industry, pop music is almost always produced by first recording each track in isolation and then mixing them and adding effects. This procedure, however, does not apply to classical music, which involves much less processing. Different instrumental parts of a classical music piece are almost always recorded together. This is why these datasets do not contain many classical ensemble pieces. 
The multi-track recordings in ENST-Drums, Abe{\ss}er et al., and EEP were recorded using multiple microphones simultaneously, hence there are no synchronization issues with the recording. However, leakage between microphones is inevitable for the first two of these datasets, and the contact microphone used in EEP alters the acoustic properties from normal near-field microphone recordings, which makes the dataset less desirable for source separation research.

Existing multi-track datasets that have dealt with the synchronization issue when recording each track in isolation are WWQ, TRIOS, Bach10, and PHENICX-Anechoic. WWQ only has one quintet piece, and the recording process has two stages. In the first stage the performers played together with separate microphones, one for each instrument. Audio leakage inevitably existed in these recordings but they served as a basis for synchronization in the second stage. In the second stage players recorded their parts in isolation while listening to a mix of the other player's recordings in the first stage through earphones. Because these players had rehearsed together and listened to their own performance (the first-stage recordings), the synchronization among individual recordings in the second stage was very accurate. In TRIOS, for each piece, a synthesized audio recording was first created for each instrument from the MIDI score. Each player then recorded his/her part in isolation while listening to the mix of the synthesized recordings of other parts synchronized with a metronome through earphones. Although the players did not rehearse together prior to the recording, the synchronization was easy to address as all the pieces have a steady tempo. In Bach10, instead of using the synthesized recordings and a metronome as the synchronization basis, each player listened to the mix of all previously recorded parts. The first player, however, determined the temporal dynamics and did not listen to anyone else, resulting in a less-than-ideal synchronization in Bach10. Due to significant variation in the tempo, a listener could easily find many places where notes were not articulated together. In fact, each piece contains several fermata signs, where notes were prolonged beyond their normal duration when the performance was held. 
For recording the dataset as annotated by PHENICX-Anechoic, a pre-recorded conducting video with a pianist playing was used to set the common timing for the instrumental players. The detailed description of the dataset creation process is presented in \cite{patynen2008anechoic}. In this paper, we arrived at the same synchronization approach as that in~\cite{patynen2008anechoic} independently. Unlike~\cite{patynen2008anechoic} where an audio-only dataset was generated, we create a multi-modal audio-visual dataset that we compare comprehensively with existing datasets and also set up performance baselines for several typical tasks on the dataset. Additionally, we also provide a quantitative assessment of alternative synchronization approaches, which has not previously been done.

\section{Approaches to Synchronization}
\label{sec:attempts}

\begin{figure}
\includegraphics[width=0.95\columnwidth]{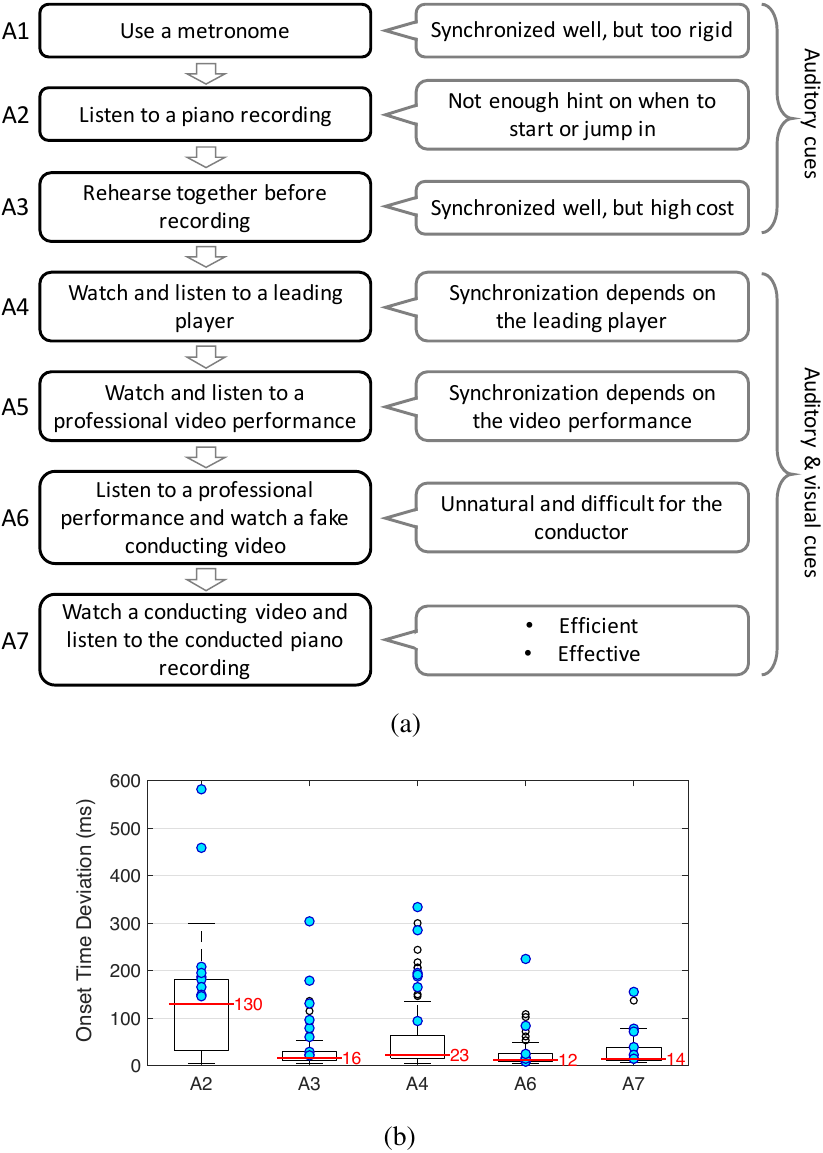}
\caption{(a): A summary of all attempts for solving the synchronization issue. (b): A quantitative evaluation on the onset time deviation among score-notated simultaneous notes of several key attempts. Red bars and text show the median values. Blue circles are the notes occurring after a rest of at least 2 beats, which are not necessarily outliers.}
\label{fig:attempts}
\end{figure}

Similar to the creation of existing multi-track datasets, the creation of URMP dataset faced the synchronization issue. This issue is even more significant because of the following seemingly conflicting goals: 1) \textit{Efficiency}. Our goal is to create a large dataset containing dozens of pieces with different instrument combinations. We also hope that each player could participate in the recording of multiple pieces. Therefore, it would be difficult and time consuming to arrange players to rehearse together before the recording for each piece, which is the approach adopted by the creation of WWQ dataset. 2) \textit{Quality}. We want the players to be as expressive as what they would be in real musical concerts. This requires them to vary the tempo and dynamics significantly throughout a piece. However, without the live interactions between players, this goal makes the synchronization more difficult. 
We tried different ways to overcome this challenge and eventually arrived at the same approach used in \cite{patynen2008anechoic} independently, which achieved both good efficiency and quality. We present our attempts here and hope that this will give some insights into the dataset creation problem. Figure \ref{fig:attempts} (a) summarizes our attempts. We also quantitatively evaluate the quality of several typical attempts by 
showing the maximal onset time deviation in Figure \ref{fig:attempts} (b), which calculates the maximal absolute time difference among the score-notated simultaneous notes from different tracks. The blue circles represent notes after a rest of at least 2 beats, which are more challenging to synchronize due to fewer temporal hints.

A1) The first approach that we tried was to pre-generate a beat sequence using an electronic metronome, and then have each player listen to the beat sequence through an earphone while recording his/her part. Different instrumental tracks were thus synchronized through the common beat sequence. 
We tested this approach on a violin-cello duet (\textit{Minuet in G major} by \textit{J. S. Bach}). 
Although the synchronization was good, we found that the performance was too rigid and did not reach our desired level of expressiveness. 

A2) In order to have better expressiveness, we replaced the beat sequence with a pre-recorded piano performance. The pianist played both instrumental parts simultaneously and varied the tempo and dynamics throughout the piece. However, when the players later followed it to record their individual parts, the synchronization was not satisfactory, as shown in Figure \ref{fig:attempts} (b). They did not get enough hints on when to start nor when to jump in after a long break, which resulted in some extreme outliers (shown by the blue circles).

A3) This attempt is inspired by WWQ's approach: players rehearsed together and used their rehearsal recordings as the basis for synchronization. We further added a conductor to the rehearsal process to improve the expressiveness by varying the tempo and dynamics. The conductor also vocalized several beats before the start of the piece to signal the players. We tested on another violin-cello piece (\textit{Melody by Schumann).} Figure \ref{fig:attempts} (b) shows that the synchronization quality is greatly improved with a median onset microtiming value of 16ms. However, this approach is time-consuming and difficult to scale to many pieces with different instrument combinations.

A4) In the following attempts, we aim at an approach that does not require joint rehearsals while keeping both the synchronization and expressiveness at an acceptable level. Similar to Bach10, we let one leading player record first and let the other(s) follow. Differently, the follower(s) not only listened to but also watched the first player's recording. We tested on the same piece as in A1 and A2, and found that this approach improved the synchronization from A2, especially at places after long rests. This improvement, reported by the players, was mainly thanks to the visual cues displayed in the first player's motion. This cue, however, would depend on the leading player and the arrangement of the piece. Overall, its synchronization quality is still much worse than A3.

A5) Building on the previous approach, we asked each player to watch and listen to a professional video performance downloaded from YouTube during the recording. Due to the availability of professional performances, we chose a different piece, \textit{The Art of Fugue No. 1} by \textit{J. S. Bach} string quartet (same for the following attempts). We tested this approach only on the violin and cello parts. Our players reported that the visual cues were not always clear, and it was challenging for them to follow the professional performance even after watching it repeatedly in advance. They were not able to complete the recording using this approach hence we could not quantitatively analyze the synchronization quality in Figure \ref{fig:attempts} (b).

\begin{figure*}
	\centering
	\includegraphics[width=\textwidth]{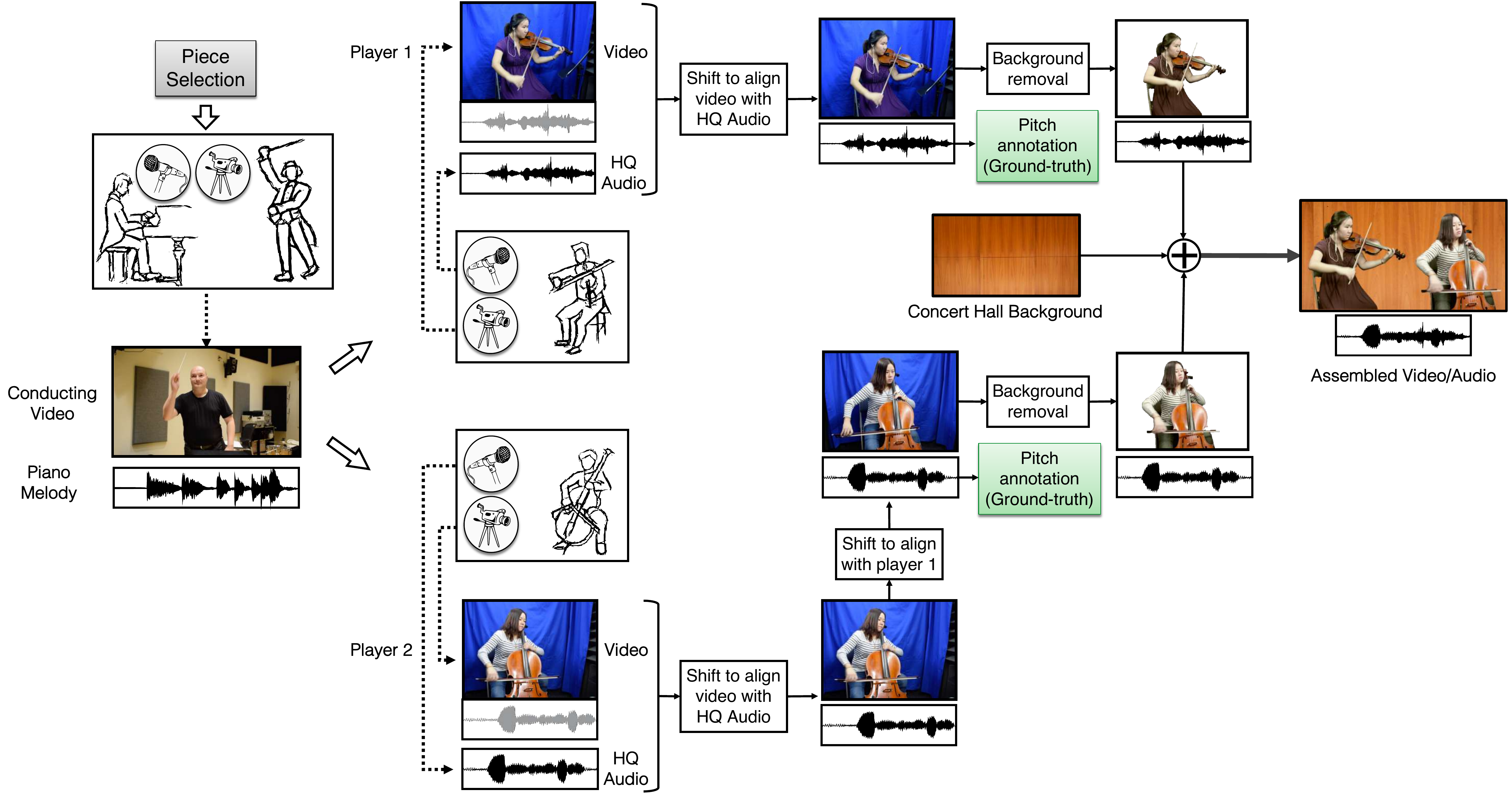}
	\caption{The general process of creating one piece (a duet in this example) of the URMP dataset.}
	\label{fig:CrePro}
\end{figure*}

A6) This attempt focused on relieving players' synchronization burden by applying a professional conductor to ``conduct'' the YouTube video used in A5. The conducting video was pre-recorded along with the played-back audio. Figure \ref{fig:attempts} (b) shows that this approach achieved synchronization quality similar to the approach with joint rehearsal (A3). However, the conductor needed to practice multiple times to memorize the temporal dynamics of the performance and behave in a timely fashion following the performance. This was very non-intuitive for conductors. In addition, it is difficult to find YouTube videos that exactly match our arrangements (e.g., instrumentation, key, and notes). Nonetheless, from this attempt, we learned that watching a conductor is more beneficial than watching other players' playing.

A7) In order to strike a balance between the burden placed on the conductor and the players, our final attempt had two key steps. In the first step, we asked the conductor to conduct a pianist performing the piece, and recorded both the conducting video and the piano audio. The conductor varied the tempo and dynamics and the pianist adjusted the performance accordingly. The conductor also gave cues to different instrumental parts in front of the camera to help players jump in after a long rest. As a second step, we asked each player to watch the conducting video as well as listen to the corresponding audio during the recording. The result from this attempt also yielded a satisfying quality. Figure \ref{fig:attempts} (b) shows a median onset deviation value of 14 ms, similar to A3 and A6. Without mandating a joint rehearsal among the players, this method simultaneously meets the requirements of quality, efficiency, and scalability. Furthermore, it is natural for the conductor, the pianist, and the players.

Because the onset times are ambiguous for some soft articulations,  the numerical evaluation of onset time deviation in Figure \ref{fig:attempts} (b) is only a limited indicator of synchronization quality. Therefore during the preliminary attempts, we also valued players' subjective evaluation. To collect players' opinions on different pieces, the attempts at synchronization were not always tested on the same piece. The synchronization difficulty of the pieces is comparable thanks to their similar tempo and expressiveness, and is representative for most of the finally selected pieces in URMP.

\section{Dataset Creation Procedure}

\label{sec:procedure}

This section explains in detail the execution of the two key steps introduced in A7 in Section \ref{sec:attempts}. It covers the entire process of dataset creation, from piece selection and musician recruitment, to recording, post-production, and ground-truth annotation. The whole process is summarized in Fig. \ref{fig:CrePro} using a duet as an example.

\subsection{Piece Selection}
\label{sec:procedure:piece}

Our criteria of piece selection were:
\begin{itemize}
\item Generality: We want to have a good coverage of polyphony, composers, and instrumentations.
\item Complexity: The pieces should be relatively simple so that all players could handle them without much practice. The duration should not be too long (ideally 1 to 2 minutes) to ease the burden of the recording process.
\item Expressiveness: To avoid rigidness, the score should allow some self interpretations by the conductor or players, such as tempo rubato, dynamic variations, and ornamentations.
\end{itemize}

Bearing these guidelines in mind, we select pieces from a sheet music website\footnote{http://www.8notes.com}, which provides thousands of simplified and rearranged musical scores of different polyphony, styles, composers, and instrumentations. We select a number of classical ensemble pieces, covering duets, trios, quartets, and quintets. Different instrumentations include string groups, woodwind groups, brass groups, and mixed groups. Percussion instruments are not included. The pieces are simple enough so most players could play them by sight-reading or after practicing for one or two times. The durations of these pieces range from 40 seconds to 4.5 minutes, and most are around 2 minutes. In most pieces, \textit{ritardando} (gradual slowing down) appears towards the end, and various expressions on notes can be applied such as \textit{trill}, \textit{mordent}, \textit{pizzicato}. This results in 44 piece arrangements, including 11 duets, 12 trios, 14 quartets, and 7 quintets. There are 28 unique pieces from 19 different composers, from which we derive different instrument arrangements and/or keys. After such adaptations, the sheet music was regenerated using Sibelius 7.5 \cite{Sibelius}. For the detailed piece list please refer to the documentation included in the dataset.

\subsection{Recruiting Musicians}
\label{sec:procedure:musician}

\begin{figure}[h]
	\centering
	\includegraphics[width=\columnwidth]{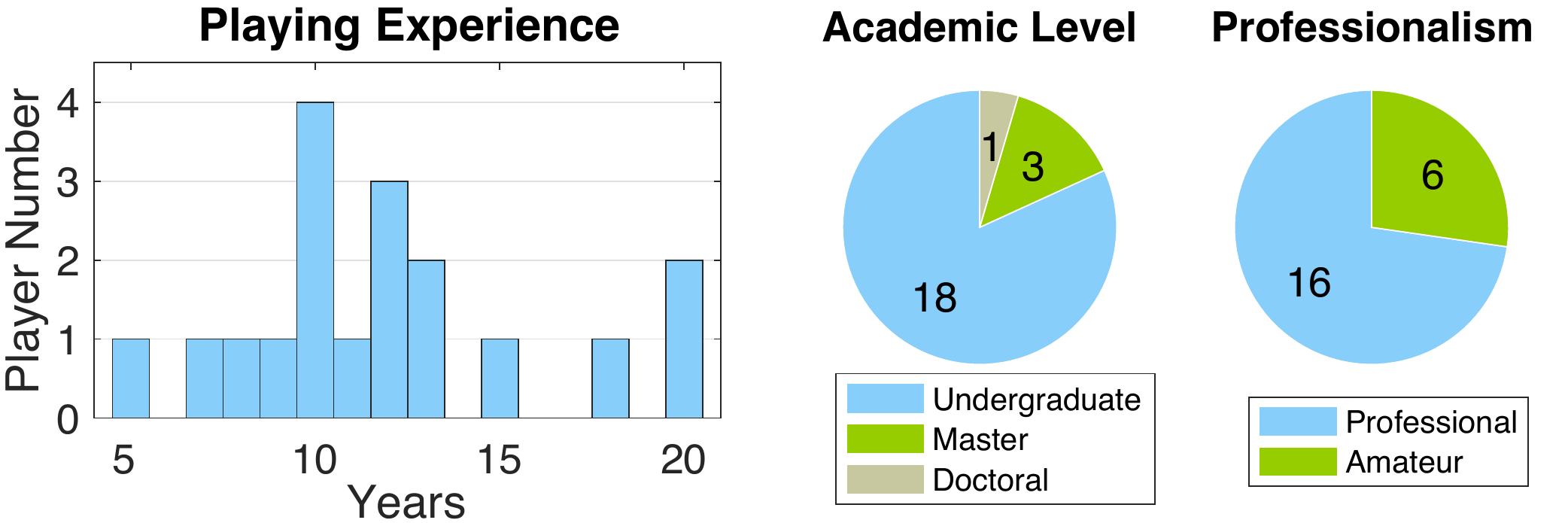}
	\caption{Demographic statistics of the total 22 musicians who recorded instrumental parts of the dataset. (The playing experience of 4 players is unknown.)}
	\label{fig:playerInfo}
\end{figure}

Creating the URMP dataset requires three kinds of musicians: a conductor, a pianist, and musicians who recorded the instrumental parts of the pieces. All of the musicians were either students from the Eastman School of Music or members of various music ensembles and orchestras from the University of Rochester. The conductor had more than 20 years of conducting experience. The pianist was a graduate student majoring in piano performance. Background statistics of the instrumental players are summarized in Figure \ref{fig:playerInfo}. In total, 22 players recorded all the instrumental parts, with each player playing only one instrument but maybe multiple tracks. All of the musicians signed a consent form and received a small monetary compensation for their participation.

\subsection{Recording Conducting Videos}
\label{sec:procedure:conducting}
For each piece, a video consisting only of a conductor conducting a pianist playing on a Yamaha grand piano was recorded to serve as the basis for the synchronization of different instrumental parts. These conducting videos were recorded in a {25}'$\times${18}' recording studio using a Nikon D5300 camera and its embedded microphone. Before recording each piece, the conductor and the pianist rehearsed several times and the conductor always started with several extra beats for the pianist (and later other players) to follow. The tempo of each piece was set by the conductor and the pianist together after considering the tempo notated in the score. All repeats within a piece were reserved for integrity. All the expression notations in the score were implemented for high expressiveness. Note that although we still need rehearsals between the conductor and the pianist for recording the conducting videos, this is much less effort than arranging joint rehearsals for all instrumental players, especially for larger ensembles and players who played in multiple pieces.

\subsection{Recording Instrumental Parts}
\label{sec:procedure:instrument}

\begin{figure}[h]
	\centering
\includegraphics[width=\columnwidth]{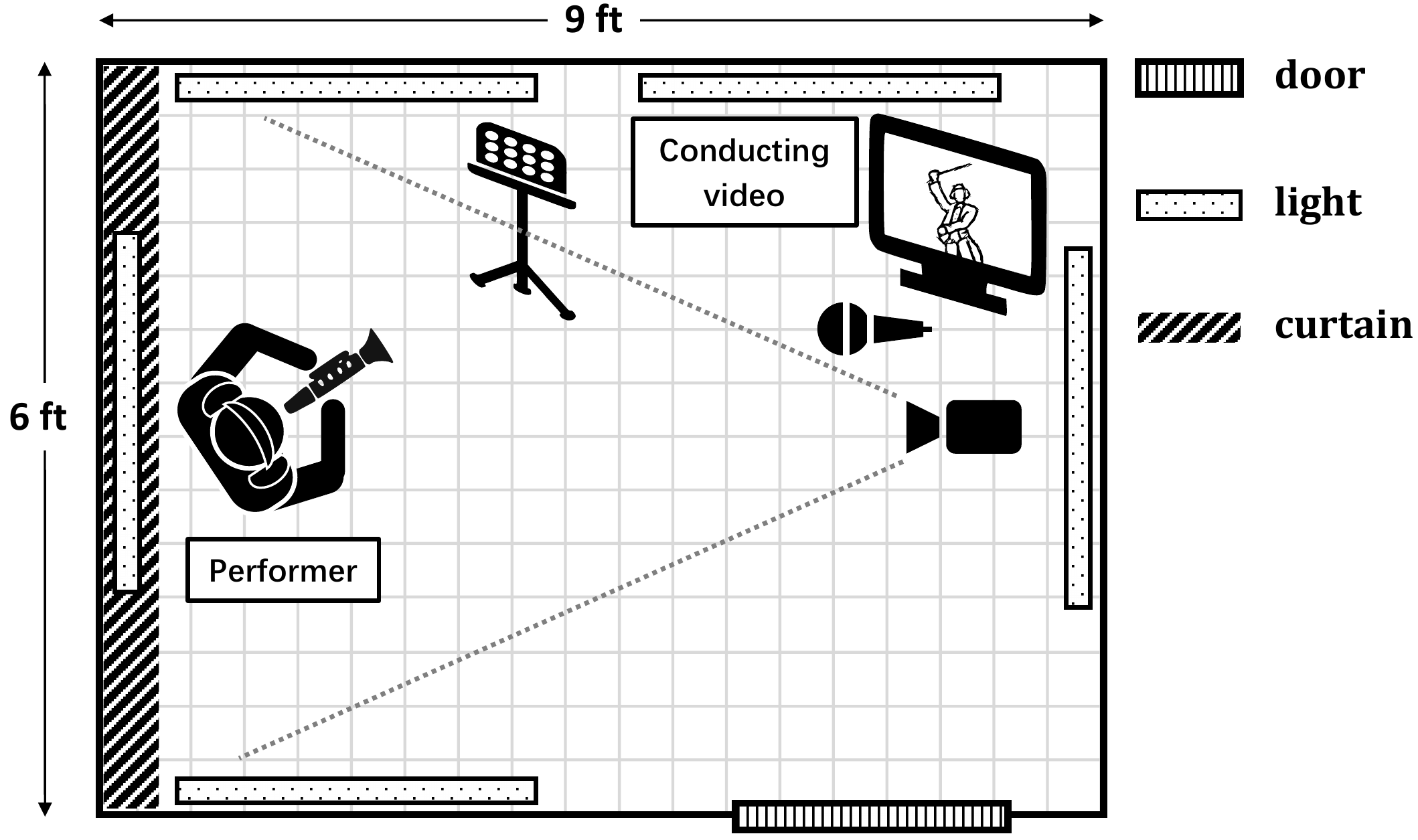}
	\caption{The floor plan of the sound booth (top-down view).}
	\label{fig_floorplan}
\end{figure}

The recording of the isolated instrumental parts was conducted in an anechoic sound booth with the floor plan shown in Figure \ref{fig_floorplan}. The wall behind the player was covered by a blue curtain and the lighting of the sound booth was through fluorescent lights affixed at wall-ceiling intersections around the room. We placed a Nikon D5300 camera on the front-right side of the player to record the video with a 1080P resolution. Since the built-in microphone of the camera did not achieve adequate audio quality, we also used an Audio-Technical AT2020 condenser microphone to record high-quality audio with a sampling rate of 48 KHz and a bit depth of 24. We connected the microphone to a laptop computer running Audacity\texttrademark~\cite{audacity} for the audio recording thereby making camera and the stand-alone microphone independently controllable. 

During the recording, the player watched the conducting video on a laptop with a 13-inch screen placed about 5 feet in front of the player. The player also listened to the audio track of the conducting video through a blue-tooth earphone with no noticeable latency. For simpler pieces, the recording was finished in one shot; while for long and difficult ones, several shots were conducted before we approved the quality of the recordings.

\subsection{Mixing and Assembling Individual Recordings}
\label{sec:procedure:mixing}
For each instrumental part, we first replaced the low-quality audio in the original video recording with the high-quality (HQ) audio recorded using the stand-alone microphone. Because the camera and the stand-alone microphone were controlled independently, the video and the high-quality audio recordings need a relative shift to ensure proper alignment, which was acccomplished automatically by using the ``synchronize clips'' function of the Final Cut Pro software \cite{FinalCutPro}.

We then assembled individual instrumental recordings. Although the individual instrumental parts of each piece were all aligned with the conducting video, the starting times of the individual recordings were not aligned with each other. We had to manually time-shift the individual recordings to align them. This was achieved by focusing on the fast sections with clear note onsets. We also manually adjusted the loudness of some tracks to achieve a better volume balance. This subjective adjustment achieved a more natural balance than objective normalization methods such as root-mean-square normalization. Then the assembled audio is the mixture (addition) of the individual high-quality audio recordings. Finally, we assembled the synchronized individual video recordings into a single ensemble recording. In the video, all players were arranged at the same level from left to right. The order of the players followed the order of score tracks.

\subsection{Video Background Replacement}
\label{sec:procedure:background}

In order to make the assembled videos look more natural and similar to live ensemble performances, we used chroma keying~\cite{Chroma:1988:key} to replace the blue curtain background with a real concert hall image. 

We use the Final Cut Pro software \cite{FinalCutPro} for video compositing. 
The blue background in the videos was unevenly lit and had players' shadow and significant textural variation. To avoid compositing artifacts due to this uneven lighting, we did color correction as a pre-processing step followed by chroma keying. By adjusting the keying and color correction parameters and by setting suitable spatial masks, we were able to get a good separation between the foreground and the background.
Once the foreground was extracted, we used a more realistic image as the background for the composite video. The background photo was captured from the Hatch Recital Hall\footnote{\url{http://www.esm.rochester.edu/concerts/halls/hatch/}} using a Nikon D5300 camera.

\subsection{Ground-truth Annotation}
\label{sec:procedure:groundtruth}

We also provide ground-truth pitch annotations for each audio track. This annotation was performed on each single audio track using the Tony melody transcription software \cite{mauch2015computer}, which implements pYIN \cite{mauch2014pYIN}, a state-of-the-art frame-wise monophonic fundamental frequency (F0) estimation algorithm. For each audio track, we generated two files: a frame-level pitch trajectory and a note sequence. The pitch trajectory was first calculated with a frame hop size of 5.8 ms, and then interpolated to 10 ms according to the standard format of ground-truth pitch trajectories in MIREX. The note sequence was extracted by the Tony software using Viterbi decoding of a hidden Markov model. The pitch of each note takes un-quantized frequencies. To guarantee a good annotation quality, we manually went through all the files introducing necessary corrections. For the frame-level pitch annotation, the annotation from the automatic tool is precise to musical cents, and we only manually corrected insertion, deletion, and octave errors. For the note-level pitch annotation, manual corrections were performed on more than half of the notes, mostly about adjusting the note onset/offset, such as splitting the wrongly merged notes. On average, the correction of each track required about half an hour. We provide the visualizations of all the annotations on the project website~\cite{URMPOverviewurl}.

\section{The Dataset}
\label{sec:dataset}

\subsection{Dataset Content}
\label{sec:dataset:content}

The URMP dataset contains audio-visual recordings and ground-truth annotations for all the 44 pieces, each of which is organized in a folder with the following content:

\begin{itemize}
\item \textbf{Score:} we provide both the MIDI score and the sheet music in PDF format. The sheet music is directly generated from the MIDI score using Sibelius 7.5 with minor adjustment (clef, key set, note spelling, etc.) for display purposes. The encoded track IDs in MIDI files are ordered following the score track order.

\item \textbf{Audio:} individual and mixed high-quality audio recordings in WAV format, with a sampling rate of 48 KHz and a bit depth of 24. The naming convention of individual tracks follows the same order as the tracks in the score.

\item \textbf{Video:} assembled video recordings in MP4 format encoded with an H264 codec. Videos have 1080P resolution (1920$\times$1080), and a frame rate of 29.97 FPS. Players are rendered horizontally, from left to right, following the same order as the tracks in the score. Additional details regarding object-level spatial resolution are provided in Section \ref{sec:dataset:resolution}.

\item \textbf{Annotation:} ground-truth frame-level pitch trajectories and note-level transcriptions of individual tracks in ASCII delimited text format. 

\end{itemize}

An overview of the dataset and a sample piece are available at~\cite{URMPOverviewurl} along with a document that lists all 44 pieces and their instrumentations. The full 12.5 GB dataset is deposited in the Dryad Digital Repository~\cite{URMP_Dryad}.

\subsection{Synchronization Quality}
\label{sec:dataset:quality}

Because maintaining the synchronization among different instrumental parts is the main challenge in creating the URMP dataset, we compare the synchronization quality of this dataset with that of Bach10 and WWQ. Both datasets have been used in the development and/or testing phases for the MIREX Multi-F0 Estimation \& Tracking task in the past. We did not include PHENICX-Anechoic because it used the same approach as URMP and its pieces are symphony pieces with many more parts than the other datasets.

\subsubsection{Quantitative Evaluation}

\begin{figure}[h]
	\centering
\includegraphics[width=\columnwidth]{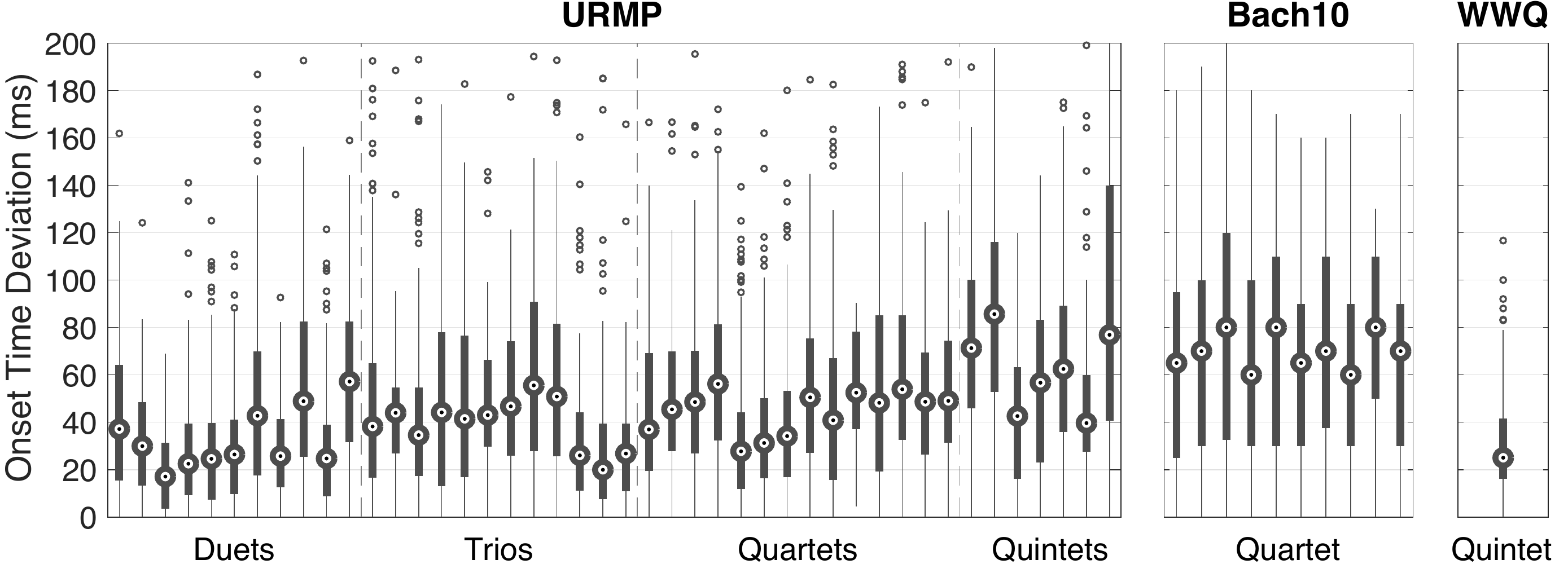}
	\caption{Synchronization quality for individual pieces in the URMP, Bach10, and WWQ dataset assessed by onset time deviation for score-notated simultaneous notes. On average, the synchronization quality is ranked as WWQ$>$URMP$>$Bach10.}
	\label{fig:synQuality}
\end{figure}

We first numerically compare the synchronization quality by calculating the onset time deviations as described in Section \ref{sec:attempts}. When the polyphony is higher than two, the maximum deviation among the score-notated simultaneous notes is calculated. Figure \ref{fig:synQuality} shows a boxplot of the maximum deviation for each piece in URMP, Bach10, and WWQ. The best synchronization quality is achieved by WWQ, where players rehearsed before recording, a methodology that does not scale to larger datasets. Also note that only a 54-second excerpt out of the 9-minute recording is publicly available and is evaluated here. This excerpt has a strong rhythmic pattern which might help the synchronization. The URMP dataset achieves the second best synchronization quality, with the maximum onset deviation being in the range of 20 to 60 ms. This deviation is larger than our preliminary evaluations in Figure \ref{fig:attempts} (b). This is because we include all of the pieces here and many of them are larger ensembles. Bach10 achieves the worst synchronization quality, showing the maximum onset deviation in the range of 60 to 80 ms. 

\subsubsection{Subjective Evaluation}

The numerical evaluation based on onset time deviations has its own limitation, considering the ambiguity of onset instances for some soft articulations. So we also conducted a subjective evaluation. We recruited 8 subjects who were students at the University of Rochester from various fields. Half of them had musical background, and none of them were familiar with these datasets. For each subject, we randomly chose pieces from these datasets to form 4 triplets, one piece from each dataset. We then asked the subjects to listen to the three pieces of each triplet and rank their synchronization quality. Table \ref{tab:SubExp} shows the ranking statistics. It can be seen that out of the 32 rankings (4 rankings per subject for 8 subjects), URMP ranks first 9 times, and ranked second 17 times. This is consistent with our quantitative evaluation.

\begin{table}[htbp]

	\centering
		\begin{tabular}{lccc}
		\hline
        \textbf{Rank} & \textbf{\#1} & \textbf{\#2} & \textbf{\#3}\\
        \hline
        \textbf{URMP} & 9 & 17 &6\\
        \textbf{WWQ \cite{bay2009evaluation}} & 22 & 9 &1\\
        \textbf{Bach10 \cite{duan2010multiple}} & 1 & 6 &25\\
        \hline
		\end{tabular}
        
\caption{Subjective ranking results of the synchronization quality of the three datasets provided by eight subjects.}
\label{tab:SubExp}

\end{table}

\subsection{Spatial Occlusion \& Resolution}
\label{sec:dataset:resolution}

In this section we analyze several aspects of the visual quality of the dataset, i.e., the spatial occlusion and resolution on Regions of Interest (ROI) where the musician-instrument interactions take place. As we mentioned in Section \ref{sec:procedure:instrument}, the videos were captured from the right-side of the players, whose locations and orientations were kept unchanged throughout the piece. So only the right-side faces are in the view without occlusions. From this camera angle, self-occlusions on the players' hands or arms vary for different instrument types:
\begin{itemize}
\item Violin/Viola: Both the right-arm bow motion and left arm is visible. The detailed fingering of the left hand is partially occluded.
\item Cello/Bass: The right-arm bow motion and left-hand fingering motion are visible. The left arm is sometimes occluded by the instrument body.
\item Woodwind: One arm is in the front and the other arm is occluded. The fingering motion of both hands are visible.
\item Brass: Only one hand contributes to the fingering and it is visible.
\end{itemize}

\begin{figure} [h]
	\centering
	\includegraphics[width=\columnwidth]{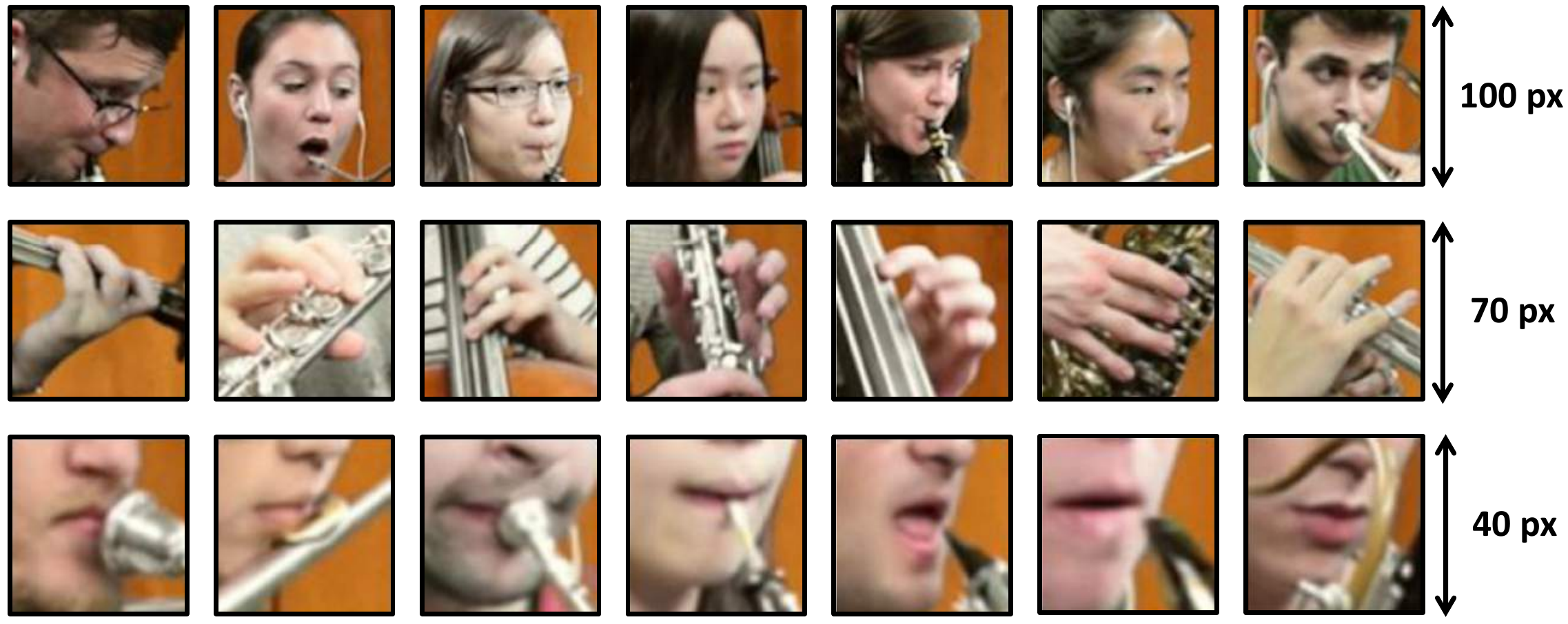}
	\caption{Spatial resolution of ROIs (face, hand, mouth) where most musician-instrument interactions take place.}
	\label{fig:spatialResolution}
\end{figure}

The spatial resolution on ROIs is a relevant parameter to know which tasks can be tackled using our dataset. Since we used a fixed camera-player distance through the whole recording process, this resolution is roughly the same for all the individual video recordings. After rescaling the players' size in the 1080P assembled video (with some resolution loss), the players' faces, hands, and mouths have the resolutions of about 100$\times$100, 70$\times$70, 40$\times$40 pixels, respectively. We sample several typical ROIs from video frames and indicate corresponding spatial resolutions in Figure \ref{fig:spatialResolution}. Note that the resolution loss from individual to assembled videos depends on the ensemble size. For example, for the same ROI, the spatial resolution of a quintet is slightly lower than that of a duet, as more players occupy the 1080P video frame.

\subsection{Limitations of the Dataset}

Although we used our resources to create a high-quality audio-visual dataset, there are still limitations we need to point out, which may prevent some potential usage. The limitations mainly exist in the visual part: the limited camera view. Throughout the whole recording process, only one camera was used, so the videos all have a single-camera view. Alternatively, stereo (or even multi-view) datasets are becoming available nowadays and can support more tasks, e.g., depth estimation, 3D reconstruction. Also, our camera view is not always optimal. For example, it is difficult to infer the pitch being played by a violinist from the finger position on the string board, even though the fingering motion is generally visible. Also, in some scenarios, important objects such as the end point of a violin bow and the head of the bass player, are outside the camera view. This is because of the limited size of the sound booth. The single-camera view limitation also makes the arrangement of players in the assembled videos less natural: players all face to the same direction, which rarely happens in real chamber music performances.

Another limitation of the assembling process is that possible occlusions between players or by the music stand were not considered. This may make the video analysis on our dataset easier than real scenarios. Also, because the instrumental parts were recorded in isolation, natural interactions among players such as eye contacts and body motion interactions do not exist in the assembled videos even though such interactions are commonly observed in real performances. Thus player interactions cannot be visually analyzed using the dataset.

There are several other minor issues that could be avoided in the future work of dataset creation. For example, the bluetooth earphone still has a short wire which resulted in irrelevant movements. The chroma keying operation during the background replacement step sometimes causes slight changes in the color of the foreground.

\section{Applications of The Dataset}
\label{sec:applications}

As the first audio-visual multi-track multi-instrument music performance dataset, URMP can support a large variety of MIR tasks, several of which are highlighted in this section. In the first part, we describe two existing MIR tasks that only require the audio modality. We run well-known algorithms on URMP and another widely used multi-track music audio dataset. This also helps benchmark URMP's audio modality with existing datasets. In the second part, we propose novel tasks that require both the audio and visual modalities of URMP. We also set up evaluation metrics and provide baseline systems. We hope that the baseline results that we provide will invite other researchers to pursue these new research directions and explore other directions with URMP.

\subsection{Existing Tasks Using Only Audio Modality}
\label{sec:applications:existing}

There are many existing MIR tasks that URMP can support, and here we only describe two tasks that take the full use of the audio modality and the associated annotations: multi-pitch analysis and score-informed source separation. For these tasks, URMP can be benchmarked with suitable existing multi-track musical audio datasets. Within the multi-track category, only the Bach10, TRIOS, WWQ, and PHENICX-Anechoic have clean individual audio tracks with required annotations. The publicly available audio recording from WWQ is too short for a systematic comparison. For TRIOS, one instrument is a piano, which makes it difficult to define the polyphony and to perform a fair comparison. Also, PHENICX-Anechoic has orchestra pieces with 8-10 instrumental parts and 10-39 individual tracks, which makes the algorithm performance not comparable for the same reason. Therefore, we just use Bach10 for a comparison with URMP.

\subsubsection{Multi-pitch Analysis}
\label{sec:applications:existing:multi}

\begin{figure} [h]
\centering
\includegraphics[width=\linewidth]{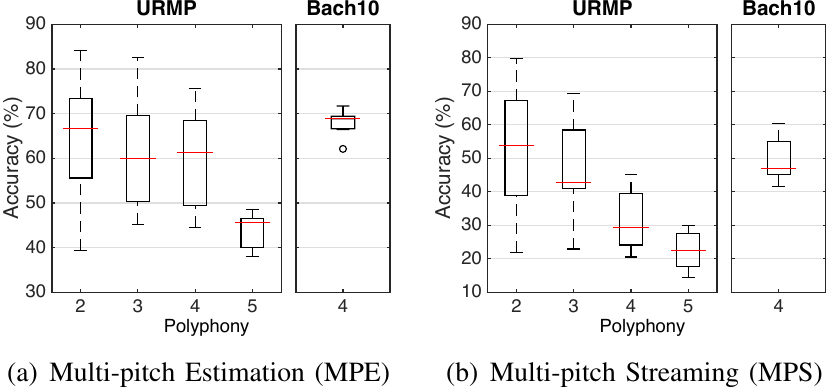}
	\caption{Comparison between URMP and Bach10 for multi-pitch analysis.}
	\label{fig:mpe}
\end{figure} 

This task consists of \emph{multi-pitch estimation (MPE)} and \emph{multi-pitch streaming (MPS)}, which are defined as estimating concurrent pitches and organizing them into temporal streams according to their sound sources, respectively. It is a fundamental task towards automatic music transcription and many other MIR applications. For MPE, we run the algorithm described in \cite{duan2010multiple}, which proposes a maximum likelihood method to model relations between the magnitude spectrum and underlying pitches. For MPS, we run the algorithm proposed in \cite{duan2014multi}, which clusters pitches into pitch streams according to their timbre and locality. Both methods are well known and have been tested on Bach10. Performance on both MPE and MPS is often measured by accuracy, which is defined as
\begin{equation}
\textnormal{Accuracy} = \frac{\#\textnormal{TP}}{\#\textnormal{TP} + \#\textnormal{FP} + \# \textnormal{FN}},
\end{equation}
where TP, FP, FN represent true positives, false positives and false negatives, respectively. They are calculated by comparing the estimated and ground-truth pitch with a tolerance of a quarter-tone~\cite{bay2009evaluation}.

The results on URMP (the first 1-min excerpt of each piece) and Bach10 are shown as boxplots in Figure \ref{fig:mpe}, where each piece constitutes one data point, and the red line in each box shows the median value. As expected, both MPE and MPS accuracies decrease when polyphony increases on URMP. When the polyphony is 4, both MPE and MPS accuracies are significantly lower than those on Bach10, suggesting that URMP is a more challenging dataset than Bach10. Indeed, URMP has a larger variety of music pieces, instrumentation, and playing techniques than Bach10, which only contains Bach chorales. Furthermore, different tracks of the same piece of URMP may use the same instrument while Bach10 always uses different instruments. This makes it more difficult to exploit the timbre cues for pitch streaming.

\subsubsection{Score-informed Source Separation}

\begin{figure}[h]
\centering
{\includegraphics[width=\columnwidth]{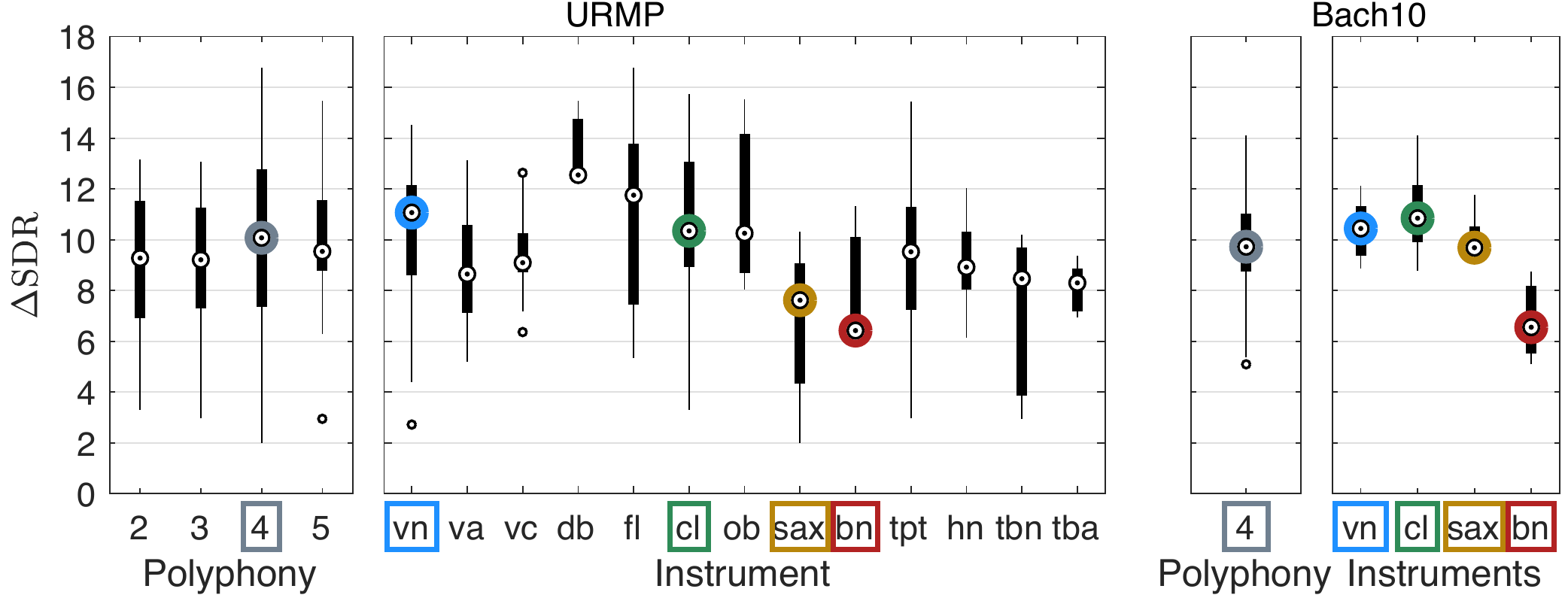}}
	\caption{Comparison between URMP and Bach10 for score-informed source separation. Colors encode overlapping categories for easier reference.}
	\label{fig:ss}
\end{figure} 

This task leverages score information to separate musical audio sources. The algorithm we use first aligns the score to the audio mixture using dynamic time warping on chroma feature sequences \cite{fujishima1999realtime}. Then audio sources are separated using harmonic masking as described in \cite{duan2011soundprism}. The quality of the separated audio sources is measured using the Signal-to-Distortion Ratio (SDR)~\cite{vincent2006performance}. We further calculate the $\Delta$SDR, which measures the improvement of SDR from the audio mixture to the separated source. Figure \ref{fig:ss} shows boxplots of the results, where each track constitutes one data point, and the circle in each box shows the median value. In contrast with the trends in Figure \ref{fig:mpe}, we can see that the performance on URMP and Bach10 is very similar, for pieces with the same polyphony (quartets) and tracks played by the same instrument. This shows that the score information helps significantly in overcoming the greater challenges posed by URMP compared with Bach10. Also, the harmonic masking method for source separation does not model timbre information; and thus underexploited the ``distinct timbre'' advantage of the Bach10 dataset.

\subsection{New Tasks Using Both Audio and Visual Modalities}

With the visual modality available, URMP not only serves for the development and evaluation of audio-based approaches, but also opens up new frontiers for MIR tasks. In this section, we propose two representative tasks that require both the audio and visual modalities. We define the tasks, set up evaluation strategies, and provide baseline results on the URMP dataset to invite the research community to pursue these new research directions.

\begin{figure}[h]
\centering
\includegraphics[width=\linewidth]{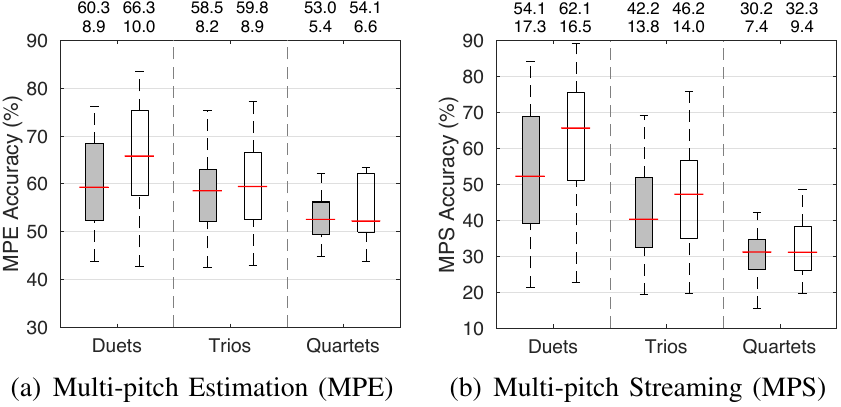}
		\caption{Comparison between the proposed visually informed method (white) and the audio-based method (gray) on the multi-pitch analysis task. For each boxplot, the mean and standard deviation values are listed above the plot. Results are reproduced from \cite{dinesh2017visually}.} \label{fig:mpe_mps}
\end{figure}

\subsubsection{Visually Informed Multi-pitch Analysis}

This is the same task as defined in Section \ref{sec:applications:existing:multi}, but here visual information is available. Visual information about the music performance can significantly help multi-pitch analysis: Observation of the fingering can directly help predict the notes being played; detection of play/non-play activity of instrument players may help estimate the instantaneous polyphony and assign pitches to correct sources. There exist several systems that utilize visual information to estimate pitches for instrument solos such as violin \cite{zhang2007visual}, piano \cite{akbari2015real}, and guitar \cite{paleari2008multimodal}, but little work has been done for other instruments or ensembles, due to the lack of datasets.

We propose to start this task with the 11 string ensembles in the URMP dataset, which provide the most pronounced motion information. Our previous work in \cite{dinesh2017visually} addresses both MPE and MPS for these pieces and can serve as a baseline for future approaches. The basic idea of this work is to model the play/non-play (P/NP) activity of each player from the visual modality and then use it to constrain audio-based pitch analysis. The P/NP activity is classified in each video frame using the bowing motion features that are calculated from optical flow estimation \cite{Secrets:2010:sun}. For MPE, the detected P/NP label provides a more accurate estimate of the instantaneous polyphony in each frame. For MPS, this label constrains the assignment of pitch estimates to sources: pitch estimates are only assigned to active players. This idea was implemented based on the same audio-based MPE/MPS algorithms as described in Section \ref{sec:applications:existing:multi}.

Figure \ref{fig:mpe_mps} compares the MPE and MPS accuracies of this method with those of the audio-based method, where each piece constitutes one data point. Note that each polyphony category is the expanded set using all the possible track combinations within each piece. An improvement between 2-12\% can be seen across the tasks and pieces.

We want to state that this baseline approach is just a preliminary attempt to address the multi-pitch analysis problem for string instruments. Much visual information such as the fingering is not exploited. In addition, reliable detection of P/NP activity for non-string instruments where motion is more subtle is also an open problem \cite{bazzica2014exploiting}.

\begin{figure}[h]
\centering
\includegraphics[width=\linewidth]{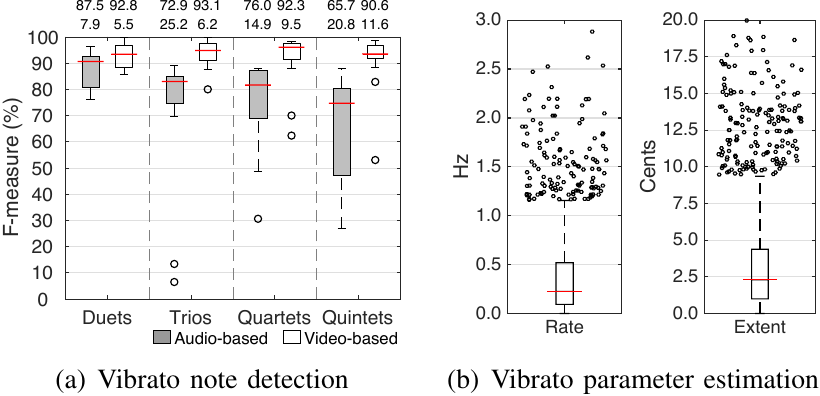}
		\caption{Video-based vibrato note detection and parameter analysis results, reproduced from \cite{li2017video}. For each boxplot, the mean and standard deviation values are listed above the plot.} \label{fig:vibrato}
\end{figure}

\subsubsection{Polyphonic Vibrato Analysis}

In music performances, vibrato is an important artistic effect that adds expressiveness and emotions by slight variations in pitch. Vibrato analysis provides basis for comparing different articulation styles, and thus has broad impact in musicological studies. It also facilitates other tasks such as melody extraction and music synthesis. However, most of the existing automatic vibrato analysis tools are audio-based with a focus on monophonic recordings. In polyphonic cases, even if the score is provided, the task is challenging due to the severe interference among sources. Existing audio-based techniques are not yet capable of this task.

The visual modality of a music performance can be very helpful for vibrato analysis. This is especially true for string instruments, where the left-hand fingers' rolling motion along the fingerboard is the direct cause of the fluctuation of pitch. Compared to the audio signals, this motion cue does not degrade as polyphony increases. This makes the polyphonic vibrato analysis task possible.

We define this task on the 19 pieces that use at most one non-string instrument in the URMP dataset. This task contains two subtasks: 1) vibrato note detection and 2) vibrato parameter (rate and extent) estimation. To obtain ground-truth annotations, we first threshold the auto-correlation value of the ground-truth pitch contour of each note to determine whether the note has vibrato or not, and then calculate the vibrato rate and extent for vibrato notes from the auto-correlation function. To evaluate vibrato note detection performance, we propose to use precision, recall, and F-measure on each track. To evaluate vibrato parameter estimation, we propose to calculate the absolute difference between the estimated value and the ground-truth value.

Our previous wok \cite{li2017video} serves as the baseline method. It tracks the left hand of each string player using the KLT tracker \cite{tomasi1991detection}, and then extracts the hand motion features by optical flow estimation \cite{Secrets:2010:sun}. The aligned score is utilized to temporally segment the raw motion features into temporal-spatial blocks at each note onset/offset time. We then train a support vector machine (SVM) to classify each block as vibrato/non-vibrato. For vibrato parameter estimation, we perform principal component analysis (PCA) on the raw motion features to get a 1D motion curve corresponding to the hand rolling motion along the fingerboard. This amplitude of the motion curve is then normalized by that of the corresponding noisy pitch contour extracted from the audio mixture using a score-informed pitch estimation method. Vibrato rate and extent are finally measured on the motion curve.

We compare this proposed video-based baseline method with an audio-based method that extracts pitch contours in a score-informed fashion on the vibrato note detection subtask. The results are shown in Figure \ref{fig:vibrato} (a), where each track constitutes one data point, and the red line in each box denotes the median value. In all of the polyphony cases, the video-based method always achieves a high F-measure (generally over 90\%), while the audio-based method degrades as the polyphony increases. We further evaluate the vibrato parameter estimation performance. Results show that our video-based baseline achieves an average error of 0.38 Hz for rate estimation and 3.47 musical cents for extent estimation. Boxplots of these errors are shown in Figure \ref{fig:vibrato} (b), where each vibrato note constitutes a data point, and the red lines denote the median values. 90\% of the errors are within 1 Hz and 10 musical cents, respectively.

Although the current task is limited to vibrato analysis, we anticipate that it can be extended to playing technique detection of string instruments in general \cite{su2014sparse}. These playing techniques may include vibrato and positioning from the left hand, as well as bowing/plucking, up-bow/down-bow, and legato/d{\'e}tach{\'e} bowing from the right hand. We hope that this current task will promote the use of multi-modal analysis techniques in musicological studies. Furthermore, we anticipate an extension of music performance analysis to non-string instruments in near future \cite{bazzica2016detecting, bazzica2017vision}.

\subsubsection{Other Emerging New Tasks}
Besides the two new tasks that we defined  above, several other emerging tasks can be developed based on the URMP dataset:
\begin{itemize}
\item Visually Informed Source Separation: Audio events (e.g., a violin note) are often associated with visual movements (e.g., a bowing motion) \cite{parekh2017motion}. Designing methods that can leverage visual information for source separation is an interesting task.
\item Audio-visual Source Association: A related problem to source separation is how to associate sound sources or their components (e.g., a note) to visual objects (e.g., a player). A restricted version of this task has been defined and explored in \cite{li2017see} and \cite{li2017audiovisual} for string instruments by modeling their bowing motion and vibrato motion, respectively. Such techniques can be used to design novel music streaming services that allow users to target sound tracks from the visual scene \cite{zhao2018sound}. 
\item Audio-visual Cross Modality Generation: By further modeling the audio-visual relations, one may design a system that can generate one modality from the other. Chen et al. \cite{chen2017deep} made the first attempt using conditional Generative Adversarial Networks (GAN) to cross-generate static audio spectrograms and instrument-playing images. Extending this task to consider temporal dependencies is an interesting direction \cite{shlizerman2017audio}.
\end{itemize}

\section{Conclusion}
\label{sec:conclusions}
In this paper, we presented the URMP dataset, a multi-modal music performance dataset that is useful for a broad range of research applications including source separation, music transcription, audio-score alignment, music performance analysis, etc. Synchronization of separately recorded individual instrumental parts while maintaining expressiveness is a key challenge in recording such a dataset and we discussed the approaches for addressing this challenge. The approach successfully adopted for URMP involved having individual instrument players watch and listen to a pre-recorded conducting video when recording their individual parts. Objective and subjective comparisons between URMP and two other widely used multi-track music performance datasets showed that the multi-track synchronization in URMP has a high quality. We highlighted how the URMP dataset supports existing MIR tasks and also defined two novel multi-modal MIR tasks by providing evaluation measures and baseline systems. We further proposed several emerging research directions that URMP can support. We anticipate that the URMP dataset will become a valuable resource for researchers in the field of music information retrieval and multimedia.

\section*{Acknowledgment}
\label{sec:acknowledgement}
We thank Sarah Rose Smith, Drs. Ming-Lun Lee and Yunn-Shan Ma for participating in our various attempts at cross-player synchronization, Andrea Cogliati for recording the conducting videos, and Prof. James Beauchamp for helping us collect datasets for comparison. We also thank the three anonymous reviewers for their thorough and constructive comments, which have greatly improved this paper. This work was partially supported by the National Science Foundation grant No. 1741472.

\ifCLASSOPTIONcaptionsoff
  \newpage
\fi



\vspace*{-0.2in}

\begin{IEEEbiography}[{\includegraphics[width=1in,height=1.25in,clip,keepaspectratio]{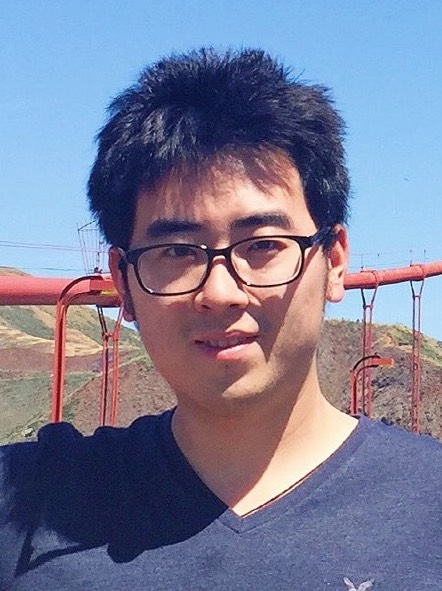}}]{Bochen Li} received his B.S. from University of Science and Technology of China in 2014. He is currently pursuing a Ph.D. degree in the Department of Electrical and Computer Engineering at the University of Rochester in the USA, under the supervision of Professor Zhiyao Duan. His research interests lie primarily in the inter-disciplinary area of audio signal processing, machine learning, and computer vision towards multimodal analysis of music performances, such as video-informed multi-pitch estimation and streaming, source separation and association, and expressive performance modeling and generation.
\end{IEEEbiography}

\vspace*{-0.25in}

\begin{IEEEbiography}[{\includegraphics[width=1in,height=1.25in,clip,keepaspectratio]{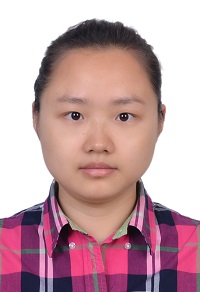}}]{Xinzhao Liu} was born in China, in 1992. She received her B.S. in Electronic Information Engineering from Dalian University of Technology, China, in 2014, and received her M.S. in Electrical and Computer Engineering from University of Rochester in 2016. She is working as a DSP engineer at Listent America Corp. 
\end{IEEEbiography}

\vspace*{-0.25in}

\begin{IEEEbiography}[{\includegraphics[width=1in,height=1.25in,clip,keepaspectratio]{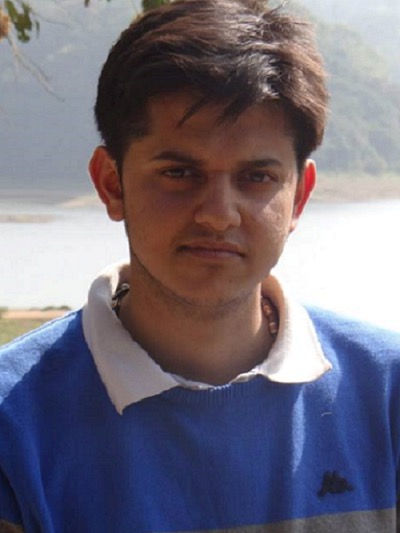}}]{Karthik Dinesh} received B.E in Electronics and Communication from National Institute of Engineering, Mysore, India in 2010 and M.Tech from Indian Institute of Technology, Kanpur, India in 2013. He is currently pursuing PhD in ECE department, University of Rochester under the supervision of Dr. Gaurav Sharma.
\end{IEEEbiography}

\vspace*{-0.25in}

\begin{IEEEbiography}[{\includegraphics[width=1in,height=1.25in,clip,keepaspectratio]{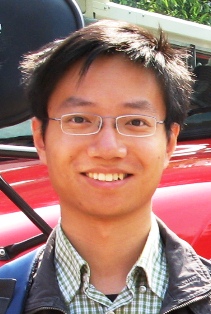}}]{Zhiyao Duan} (S'09, M'13), is an assistant professor in the Department of Electrical and Computer Engineering at the University of Rochester, where he directs the Audio Information Research (AIR) laboratory. He also holds a secondary appointment in the Department of Computer Science and he is affiliated with the Goergen Institute for Data Science. He received his B.S. in Automation and M.S. in Control Science and Engineering from Tsinghua University, China, in 2004 and 2008, respectively, and received his Ph.D. in Computer Science from Northwestern University in 2013. His research interest is in the broad area of computer audition, i.e. designing computational systems that are capable of understanding sounds, including music, speech, and environmental sounds.
\end{IEEEbiography}

\vspace*{-0.25in}

\begin{IEEEbiography}[{\includegraphics[width=1in,height=1.25in,clip,keepaspectratio]{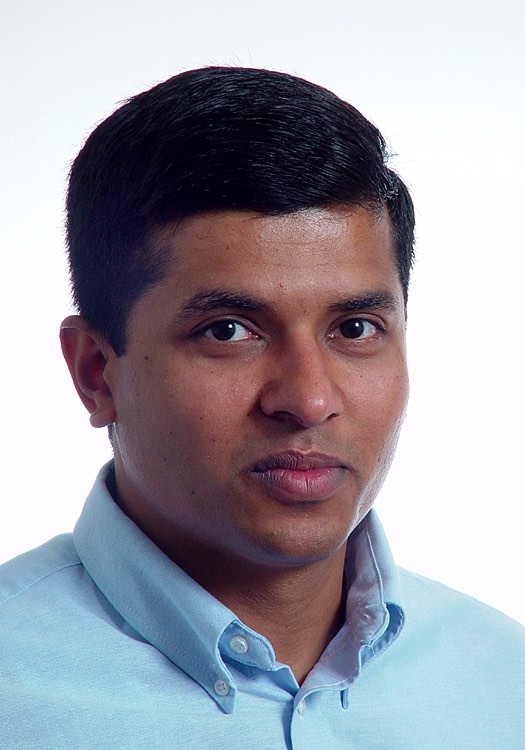}}]
{Gaurav~Sharma} (S’'88--M’'96--SM’'00--F’'13) is a professor at the University
of Rochester in the Department of Electrical and Computer Engineering,
in the Department of Computer Science and in the Department of
Biostatistics and Computational Biology. From 2008--2010, he served as
the Director for the Center for Emerging and Innovative Sciences
(CEIS), a New York state funded center for promoting joint
university-industry research and technology development, which is
housed at the University of Rochester. He received the BE degree in
Electronics and Communication Engineering from Indian Institute of
Technology Roorkee (formerly Univ. of Roorkee), India in 1990; the ME
degree in Electrical Communication Engineering from the Indian
Institute of Science, Bangalore, India in 1992; and the MS degree in
Applied Mathematics and PhD degree in Electrical and Computer
Engineering from North Carolina State University, Raleigh in 1995 and
1996, respectively. From Aug. 1996 through Aug. 2003, he was with
Xerox Research and Technology, in Webster, NY, initially as a Member
of Research Staff and subsequently at the position of Principal
Scientist.

Dr. Sharma's research interests include data analytics, image
processing and computer vision, color science and imaging, media
security, and distributed signal processing. He is the editor of the
``Color Imaging Handbook'', published by CRC press in 2003. He is a
fellow of the IEEE, of SPIE, and of the Society of Imaging Science and
Technology (IS\&T) and a member of Sigma Xi. He served as a Technical
Program Chair for the 2016 and 2012 editions of the IEEE International
Conference on Image Processing (ICIP), as the Symposium Chair for the
2013 SPIE/IS\&T Electronic Imaging symposium, as the 2010-2011 Chair
IEEE Signal Processing Society's Image Video and Multi-dimensional
Signal Processing (IVMSP) technical committee, the 2007 chair for the
Rochester section of the IEEE and the 2003 chair for the Rochester
chapter of the IEEE Signal Processing Society. Dr. Sharma is the
Editor-in-Chief (EIC) for the IEEE Transactions on Image Processing
(TIP). From 2011 through 2015, he served as the EIC for the Journal of
Electronic Imaging (JEI) and in the past has served as an associate
editor for JEI, IEEE TIP, and the IEEE Transactions on Information
Forensics and Security.
\end{IEEEbiography}




\end{document}